\documentclass[aps,prl,twocolumn,showpacs,superscriptaddress,showkeys]{revtex4-1}

\usepackage{amssymb}
\usepackage{amsmath}
\usepackage{amsfonts}
\usepackage{graphicx}
\usepackage{color}
\usepackage{xspace}
\usepackage{ulem}
\usepackage{mathtools}
\usepackage{hhline}
\usepackage{float}
\usepackage{subfigure}
\usepackage{caption}
\usepackage{subcaption}
\usepackage{subfloat}

\newcommand{\AddrCoimbra}{Univ Coimbra, Faculdade de Ci\^encias e Tecnologia da Universidade de Coimbra and CFisUC, Rua Larga, 3004-516 Coimbra, Portugal}

\begin{document}

\title{Evaporation of a Kerr-black-bounce by emission of scalar particles}

\author{Marco Calz\`a}    \email{mc@student.uc.pt}\affiliation{\AddrCoimbra}

\date{\today}

\begin{abstract}
We study a regular rotating black hole evaporating under the Hawking emission of a single scalar field. The black hole is described by the Kerr-black-bounce metric with a nearly extremal regularizing parameter $\ell=0.99r_+$. We compare the results with a Kerr black hole evaporating under the same conditions. Firstly, we compute the gray-body factors and show that the Kerr-black-bounce evolves towards a non-Schwartzchild-like asymptotic state with $a_* \sim 0.47$, differently from a Kerr black hole whose asymptotic spin would be $a_* \sim 0.555$. We show that this result depends on the combined contributions of the changes in the gray-body factors and in the surface gravity introduced by the regularizing parameter. 
We also discuss how the surface gravity affects the temperature and the primary emissivity and decreases those quantities with respect to the Kerr black hole. Consequently, the regular black hole has a longer lifetime. Finally, we briefly comment on the possibility of investigating the beyond-the-horizon structure of a black hole by exploiting its Hawking emission.
\end{abstract}

\maketitle

\section{Introduction}

General Relativity (GR) has been tested for more than one century providing outstanding results in describing the Solar System and the Universe. Despite GR's successes, its lack in addressing many open problems remains and propels the idea that it may not be the ultimate theory of gravity.
The origin of the cosmic acceleration and the nature of the dark contents of the Universe have been extensively studied using modified theories of gravity \cite{Tsujikawa:2010zza}. In recent years, the detection of Gravitational Waves (GW) from Black Hole (BH) coalescence \cite{LIGOScientific:2016aoc} by the LIGO/Virgo Collaboration and the direct observation of the BH shadows at the center of the Milky Way \cite{EventHorizonTelescope:2022xnr} and M87 \cite{EventHorizonTelescope:2019dse}  by the Event Horizon Telescope (EHT) provided a new test bench capable of probing GR robustness in a strong-field regime \cite{Banerjee:2019nnj,Mark:2017dnq,Bueno:2017hyj,Lima:2021las}.
The existence of singularities, namely portions of spacetime with an infinite curvature, is a hint that the classical framework of GR should break down or, at least, be incomplete at high energies. It is a commonly accepted the idea that singularities just reveal our lack of knowledge in the high energy regime and the related problem may be cured by a quantum theory of gravity.
Unfortunately, a theory of quantum gravity is not yet developed despite the several proposals. Nevertheless, it is still possible to gain intuition by postulating the existence of regularized spacetimes inspired by quantum gravity arguments and studying whether these new metrics give rise to new signatures or modify preexisting characteristics.
Since the 90s these motivations have led the research of regularized metrics mimicking the behavior of BH solutions \cite{Visser:1992qh}. Furthermore, in light of the new available high-energy regime tests, the field gained even more traction, and many studies about quasi-normal modes, superradiant regimes, and instabilities are regularly announced \cite{Simpson:2021dyo,Simpson:2021zfl,Daghigh:2020mog,Ovgun:2018gwt,Sebastiani:2022wbz,Rincon:2018sgd,Panotopoulos:2019qjk,Panotopoulos:2019gtn,Panotopoulos:2020mii,Gonzalez:2021vwp,Okyay:2021nnh,Gonzalez:2017shu,Chinaglia:2017uqd,Colleaux:2017ibe,Bertipagani:2020awe,Anabalon:2012tu,Babichev:2020qpr}.\\
An interesting regular metric was proposed in \cite{Simpson:2018tsi} and further analyzed in \cite{Lobo:2020ffi}.
This spacetime configuration, known as black-bounce, interpolates between the standard and regularized Schwarzschild BH and the Morris-Thorne traversable wormhole by introducing an additional parameter, $\ell$. The black-bounce metric caused a fervent activity leading to many studies of its characteristics \cite{Ovgun:2020yuv,Nascimento:2020ime,Lobo:2020kxn,Churilova:2019cyt,Tsukamoto:2020bjm,Bronnikov:2021liv,Lima:2020auu,Barrientos:2022avi} and was recently extended in order to account for rotation \cite{Lima:2021las,Mazza:2021rgq}, and afterward rotation and charge \cite{Franzin:2021vnj}. The Kerr-black-bounce and Kerr-Newman-black-bounce have also been the subject of many studies \cite{Lima:2021las,Franzin:2022iai,Yang:2022xxh,Guerrero:2021ues,Xu:2021lff,Tsukamoto:2022vkt,Ou:2021efv}.\\
The main motivation of this paper is to further enlarge the analysis of the Kerr-black-bounce  characteristics by considering its dynamical evolution due to Hawking evaporation driven by a singular scalar field. Such characteristics are certainly irrelevant for BHs of the size we measure today but may become a powerful and handy tool in light of the possible future measurement of primordial BHs.\\
The lesson of this study is that under the same conditions, a Kerr-black-bounce is characterized by a dynamic behavior that differs with respect to its singular counterpart. This work points out that tracking the evolution of a black hole spin and its spectrum will provide information on the spacetime structure.\\
This paper is organized as follows. Section II contains a brief review of the Kerr-black-bounce. Section III shows the equation governing the scalar perturbation of the Kerr-black-bounce metric, the evolution of the metric under a single scalar emission, and the numerical method used for calculating the Gray-Body Factors (GBFs). In section IV the results are presented. Section V provides a summary in which future perspectives are considered.
\\
We use units of $G = c = \hbar = 1$.

\section{Kerr-black-bounce metric}

In this section we briefly review the Kerr-black bounce metric \cite{Mazza:2021rgq}:
\begin{equation}\label{metric1}
\begin{split}
    ds^2 =&- \left(1- \frac{2M \sqrt{ \tilde{r}^2 + \ell^2}}{\Sigma}\right) dt^2 + \frac{\Sigma}{\Delta} d\tilde{r}^2 + \Sigma d\theta^2 \\ 
    &+\frac{A\sin^2 \theta}{\Sigma} d\phi^2 - \frac{4Ma\sqrt{\tilde{r}^2+ \ell^2}\sin^2\theta }{\Sigma}dt d\phi ,
\end{split}
\end{equation}
where $M$, $a$, and $\ell$ are the parameters describing mass, spin, and the regularizing parameter of the metric, while
\begin{equation}
\begin{split}
   &\Sigma = \tilde{r}^2 +\ell^2+a^2 \cos^2 \theta, \\ &\Delta= \tilde{r}^2 +\ell^2 +a^2 - 2M\sqrt{\tilde{r}^2 \ell^2} ,\\  &A=( \tilde{r}^2 +\ell^2+a^2)^2 - \Delta a ^2 \sin^2 \theta .
   \end{split}
\end{equation}
This is a generalization of the static and spherically symmetric metric proposed by Simpson and Visser \cite{Simpson:2018tsi,Lobo:2020ffi,Simpson:2019cer}. It is a stationary, axially symmetric metric which, by introducing a positive parameter, $a<M$, describes the angular momentum of the black-bounce. This line element was recently further extended in order to describe a charged spacetime \cite{Franzin:2021vnj}.\\
When the positive regularizing parameter $\ell \rightarrow 0$,  the Kerr-black-bounce metric reduces to the singular Kerr solution, while for $\ell\neq0$ the spacetime is regular and possesses a wormhole throat at $\tilde{r}=0$. A coordinate singularity interpreted as an event horizon is present when $\Delta=0$, or, equivalently, when
\begin{equation}
   \tilde{r}_\pm = \sqrt{r^2_\pm -\ell^2},
\end{equation}
where $r_\pm = M \pm \sqrt{M^2-a^2}$.\\
Depending on the values of the regularizing parameter $\ell$, the metric (\ref{metric1}) describes a wormhole for $\ell>r_+$, for which no coordinate singularities are present on the manifold. If $\ell<r_+$ the metric (\ref{metric1}) describes a BH which may have one or two coordinate singularities depending on $r_- < \ell<r_+$ or $\ell<r_-$, respectively. Finally, when $\ell=r_+$ the throat and the event horizon coincide.\\
To better visualize this interplay, it is convenient to define a new radial coordinate $r=\sqrt{\tilde{r}^2+ \ell^2}$ and pass from an extrinsic description of the manifold to an intrinsic one.
It is easy to notice that $r\neq0$ for all $\ell\neq0$. In particular, the minimum value of $r$ corresponds to the minimal radius of the throat. The coordinate $r$ measures the distance from the center of the object.
Given this new coordinate, the metric reads
\begin{equation}\label{metric2}
    \begin{split}
    ds^2=&- \left(1- \frac{2Mr^2}{\Sigma}\right) dt^2 + \frac{\Sigma}{\delta\Delta} dr^2 \\ &+ \Sigma d\theta^2 +\frac{A\sin^2 \theta}{\Sigma} d\phi^2 - \frac{4Mar\sin^2\theta }{\Sigma}dt d\phi ,
    \end{split}
\end{equation}
and
\begin{equation}
\begin{split}
   &\Sigma =r^2 +a^2 \cos^2 \theta, \\ &\Delta= r^2 +a^2 - 2Mr , \\ &A=(r^2 + a^2 )^2 - \Delta a ^2 \sin^2 \theta, \\  &\delta=1-\frac{\ell^2}{r^2} .
\end{split}
\end{equation}
If $\ell\neq0$, the curvature singularity at $r=0$ is always prevented by the wormhole throat.
When $\ell>r_+$ the wormhole throat is located at a larger radial value than the coordinate singularity of the event horizon. In this way, the presence of the horizon is prevented by the regular finite surface of the wormhole throat.
If $0\neq\ell<r_+$, the throat of the wormhole is enclosed by the event horizon and the metric describes a BH.\\
In the following part of this paper, we focus on regular BHs avoiding coordinates singularities and inner horizons. The absence of the inner horizon is a desirable feature since it might avoid the problems related to mass inflation. Moreover, this choice allows a nearly maximal value of $\ell$ for which the metric (\ref{metric1}) mostly differs from the Kerr BH and still describes a BH.\\
It has to be noticed that the metric (\ref{metric1}) or, equivalently, (\ref{metric2}), is inspired by the reasonable quantum gravity argument of avoiding singularities and other pathology, and it is not a vacuum solution of GR.

\section{Scalar perturbations and evolution}

In this section, we derive the equation describing the scalar massless perturbations of the metric (\ref{metric1}) and discuss the appropriate boundary conditions. The massless Klein-Gordon equation $\nabla^{\mu} \nabla_{\mu} \Phi=0$ in curved spacetime reads
\begin{equation}\label{KG curve}
    \frac{1}{\sqrt{-g}} \partial_{\mu} (\sqrt{-g} g^{\mu \nu} \partial_{\nu}) \Phi =0.
\end{equation}
Taking into account the decomposition $\Phi = R_{l m}(r) S_{l m}(\theta) e^{i m \phi } e^{-i \omega t }$ where $\omega$ is the perturbation frequency, $m$ is the azimutal quantum number, (\ref{KG curve}) separates into an angular equation
\begin{equation}\label{ang}
    \frac{1}{\sin \theta} \frac{d}{d\theta}\left(\sin \theta \frac{d}{d\theta} S_{l m}\right)+ \left(a^2 \omega^2 \cos^2 \theta + A_{l m} - \frac{m^2}{\sin^2 \theta}\right)S_{l m}=0
\end{equation}
describing the spheroidal harmonics equation where $A_{l m}$ are its eigenvalues, and a radial equation
\begin{equation}\label{rad}
   \sqrt{\delta} \frac{d}{dr} \left( \sqrt{\delta} \Delta \frac{d R_{l m}}{dr} \right) + \left( \frac{K^2}{\Delta} + 2 a m \omega - a^2 \omega^2 - A_{l m } \right)R_{l m}=0,
\end{equation}
where $K=(r^2 + a^2) \omega - a m$.\\
The angular equation (\ref{ang}) is the spin-less case of the well-studied spin-weighted spheroidal harmonics equation \cite{crossman,Seidel:1988ue,Berti:2005gp}. To leading order $A_{l m}=-l(l+1)+ \mathcal{O}(a \omega)$ and the $\mathcal{O}(a \omega)$ correction can be expressed as a power series in $a \omega << 1$, which are given in \cite{Berti:2005gp}. Besides, for our purposes, it is worth studying the radial equation (\ref{rad}) in two limits, near the horizon, and at spatial infinity.
If the regularizing parameter satisfies $\ell< r_+$ and the Kerr-black bounce metric (\ref{metric1}) describes a regular BH, then the near-to-the-horizon solution reads \cite{Franzin:2022iai},
\begin{equation}
\begin{split}
    &R(r) \sim (r-r_+)^{\pm i \sigma}, \\ &\sigma= \frac{a m - 2 M \omega r_+}{\gamma (r_+-r_-)},\\  &\gamma = \sqrt{1-\frac{\ell^2}{r_+^2}},
\end{split}
\end{equation}
while the far-away solution simply reads
\begin{equation}
    R(r)\sim \frac{1}{r} e^{\pm i \omega r}.
\end{equation}
To study a BH described by (\ref{metric1}), evolving by the sole emission of scalar particles due to Hawking radiation, it is necessary to set up a scattering-like problem and take into account in-going and out-going boundary conditions at infinity, while at the event horizon, one must consider pure absorption.
Those asymptotic solutions and the conservation of energy fluxes, both at the horizon and at infinity, allow one to calculate the GBF or transmission coefficient, defined as 
\begin{equation}
    T= \frac{{dE_{hole}}/{dt}}{{dE_{in}}/{dt}}.
\end{equation}
The GBFs depend on the modes, and, at a constant $\ell$, are functions of both the BH spin parameter and frequency of the perturbation, $T= { T ^l_m }(a , \omega)$. The GBFs emerge as a consequence of a geometrical potential in equation (\ref{rad}) which, acting as a barrier, partially shields the Hawking radiation from being totally emitted. This way the radiation emerging from the BH is not the one of a black body.
The field quanta have energy and spin and their emission comes at the expense of both the BH mass and angular momentum. Following the path outlined in \cite{Page:1976ki}, the rates of mass and angular momentum loss are called $f$ and $g$, respectively, and they read
\begin{equation}\label{f_g}
\begin{pmatrix}
f\\
g
\end{pmatrix}=\sum_{i,l,m}\frac{1}{2\pi} \int_0^{\infty}dx \frac{T_{i,l,m}}{e^{2\pi k/\kappa}- 1}
\begin{pmatrix}
x\\
ma_*^{-1}
\end{pmatrix},
\end{equation}
where the sum is taken over all particle species $i$, and $l$, $m$ are the usual angular momentum quantum numbers. Here $x=\omega M$, $k=\omega-m\Omega$ and
\begin{equation} \label{surface}
   \kappa=\sqrt{\frac{r_H^2}{r_H^2 + \ell^2}}\sqrt{1-a_*^2}/2r_+ 
\end{equation}
is the surface gravity of the BH \cite{Visser:1992qh,Franzin:2021vnj}. Since the choice of analyzing a singular Kerr BH $\ell=0$ and the regular Kerr-black-bounce BH having $\ell=0.99 r_+$,
the pre-factor $\sqrt{{r_H^2}/(r_H^2 + \ell^2)}$ takes just two values accordingly.
To determine whether a BH spins up or down during its evolution it is necessary to calculate the mass to angular momentum loss rates. For this reason, one defines
\begin{equation}
    h=\frac{g}{f}-2.
\end{equation}
A root of the function $h$, $\tilde{a}_*$, for which $h'(\tilde{a}_*)>0$, represents a stable state towards which the BH evolves while evaporating.
To investigate the temporal evolution of angular momentum and mass it we followed the path outlined in \cite{Page:1976ki,Page:1976df,Page:1977um} and later in \cite{Chambers:1997ax,Taylor:1998dk} defining
\begin{equation}
    y=-\ln{a},
\end{equation}
\begin{equation}
    z=-\ln{M/M_i},
\end{equation}
and 
\begin{equation}
    \tau=-M^{-3}_i t,
\end{equation}
where $M_i$ is the initial mass of the BH.
The evolution is then fully determined by the differential equations
\begin{equation}
   \frac{d}{dy} z=\frac{1}{h},
\end{equation}
\begin{equation}
   \frac{d}{dy} \tau=\frac{e^{-3z(y)}}{h f},
\end{equation}
and the initial conditions $z(t=0)=0$ and $\tau(t=0)=0$.
To estimate the primary spectrum of scalar particles we used the well-known formula \cite{Page:1976ki,Page:1976df,MacGibbon:2015mya,MacGibbon:2010nt,MacGibbon:2007yq,MacGibbon:1990zk,MacGibbon:1991tj,MacGibbon:1991vc,Halzen:1991uw,Halzen:1990ip,Ukwatta:2009xk,Ukwatta:2015iba,Calza:2021czr}:
\begin{equation}\label{emissss}
\frac{d^2N}{dt dE}=\frac{1}{2\pi}\sum_{l,m}\frac{T_{l,m}(\omega)}{e^{2\pi k/\kappa}-1}~.
\end{equation}

\subsection{Numerical method}

An explicit analytical calculation of the GBFs is possible only under stringent approximations and numerical methods are usually required to evaluate them. 
We implemented a code based on the so-called shooting method which has been applied to solve similar problems, for example in \cite{Rosa:2012uz,Rosa:2016bli}, and allows for the calculation of the  GBFs with good accuracy.\\
The first step is to rewrite Eq.(\ref{rad}) in terms of the re-scaled coordinate
\begin{equation}\label{varx}
    x=\frac{r-r_+}{r_+},
\end{equation}
such that:
\begin{equation}\label{Teuk2}
\begin{split}
    &\delta x^2(x+\tau)^2 \partial_x^2 R(x) \\ &+2x(x+\tau)\left( \frac{1}{2}(2x+\tau \delta + \frac{x(x+\tau)}{x+1}(1-\delta)) \right) \partial_x R(x) \\ &+ V(\omega,x)R(x)=0,
\end{split}
\end{equation}
where
\begin{equation}
    V(\omega,x)=\mathcal{K}^2-x(x+\tau)(A_{l m }+a^2 \omega^2-2am\omega ),
\end{equation}
with $\tau=(r_+ - r_-)/{r_+}$,
$\mathcal{K}=\varpi+x(x+2)\bar{\omega}$,
$\varpi= (2-\tau)(\bar{\omega}-m \bar{\Omega}_+)$,
where $\bar{\omega}=r_+ \omega$,
$\bar{\Omega}_+ = r_+ {\Omega_+}$
and $\Omega_+={a}/{2Mr_+}$.\\
Setting purely in-going boundary condition near the horizon, the solutions of Eq. (\ref{Teuk2}) can be expressed in the form of the Taylor expansion \cite{Rosa:2012uz,Rosa:2016bli} of the form
\begin{equation}\label{near}
    R(x)= x^{- i \varpi/(\gamma \tau)} \sum_{n=0}^\infty a_n x^n.
\end{equation}
The coefficients $a_n$ can be determined by substituting (\ref{near}) in (\ref{Teuk2}) and solving iteratively the algebraic equations.\\
The near horizon solution is used to set the boundary conditions and numerically integrate the radial equation up to large distances, where the general form of the solution takes the form:
\begin{equation}
    R(x) \rightarrow \frac{ Y^{l m }_{in}}{r_+} \frac{e^{-i \bar \omega x}}{x} + \frac{ Y^{l m }_{out}}{r_+} \frac{e^{i \bar \omega x}}{x}.
\end{equation}
It is then possible to extract the coefficient $Y^{l m}_{in}(\omega)$ in order to evaluate the GBF.
The normalization of the scattering problem is set by requiring $a_0=1$, this way GBFs read
\begin{align}
     &T^{l m}(\omega)= \frac{\varpi}{  \bar\omega }  | Y^{l m}_{in}(\omega)|^{-2}.
\end{align}
With this method, we computed the GBFs of a scalar perturbation on a regular BH described by the Kerr-black-bounce metric having a nearly extremal regularizing parameter ($\ell=0.99 r_+$). Different values for the spin parameter of the BH spanning from $a_*=0$ to $a_*=0.99$ are considered and the GBFs are calculated the up to $l=4$. This last choice is motivated by the definition of the functions $f$ and $g$. In fact the higher is the $l$ mode, the higher the energies at which the gray-body factor is non-vanishing, so the $l$ mode contribution in \ref{f_g} is smaller with respect to the $l-1$ mode.

\section{Results}

Let us compare the scalar perturbations of the Kerr BH and the ones of the nearly extremal Kerr-black-bounce 
\begin{figure}[H]
    \centering{\includegraphics[width=0.4\textwidth]{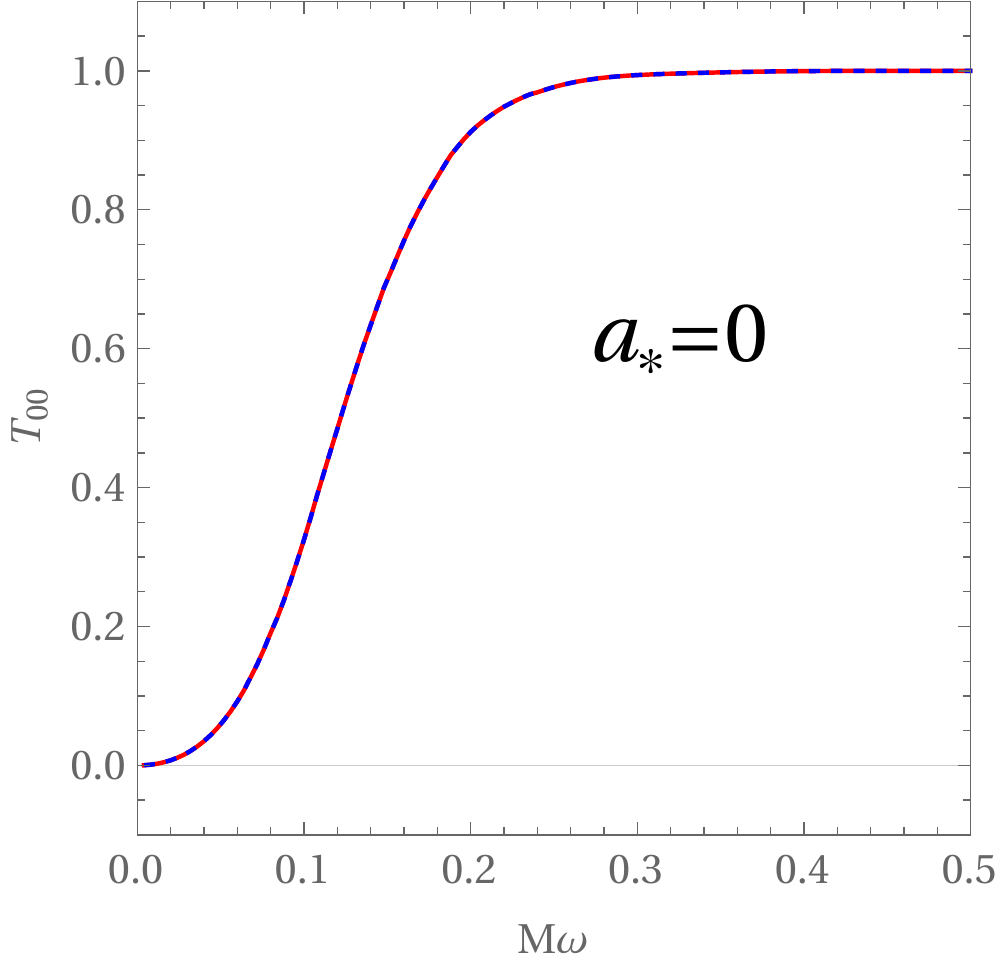}} 
    \caption{GBFs of the mode $l=m=0$ of a BH rotating at $a_*=0$ in the case $\ell=0.99r_+$ (solid red line) and $\ell=0$ (blue dashed line).}
    \label{l=m=0}
\end{figure}
\noindent BH. Those BHs share many characteristics such as the presence of a superradiant regime and a non-null asymptotic value of the spin parameter $a_*$. Nevertheless, for the two different metrics, the phenomenology changes and it is of great interest to analyze those differences.\\
The GBFs of the modes $l=m=0$ are identical (as shown in Fig. \ref{l=m=0} for the non-rotating cases) and this equality is independent of the BH spin considered.
\begin{figure}[H]
    \centering{\includegraphics[width=0.4\textwidth]{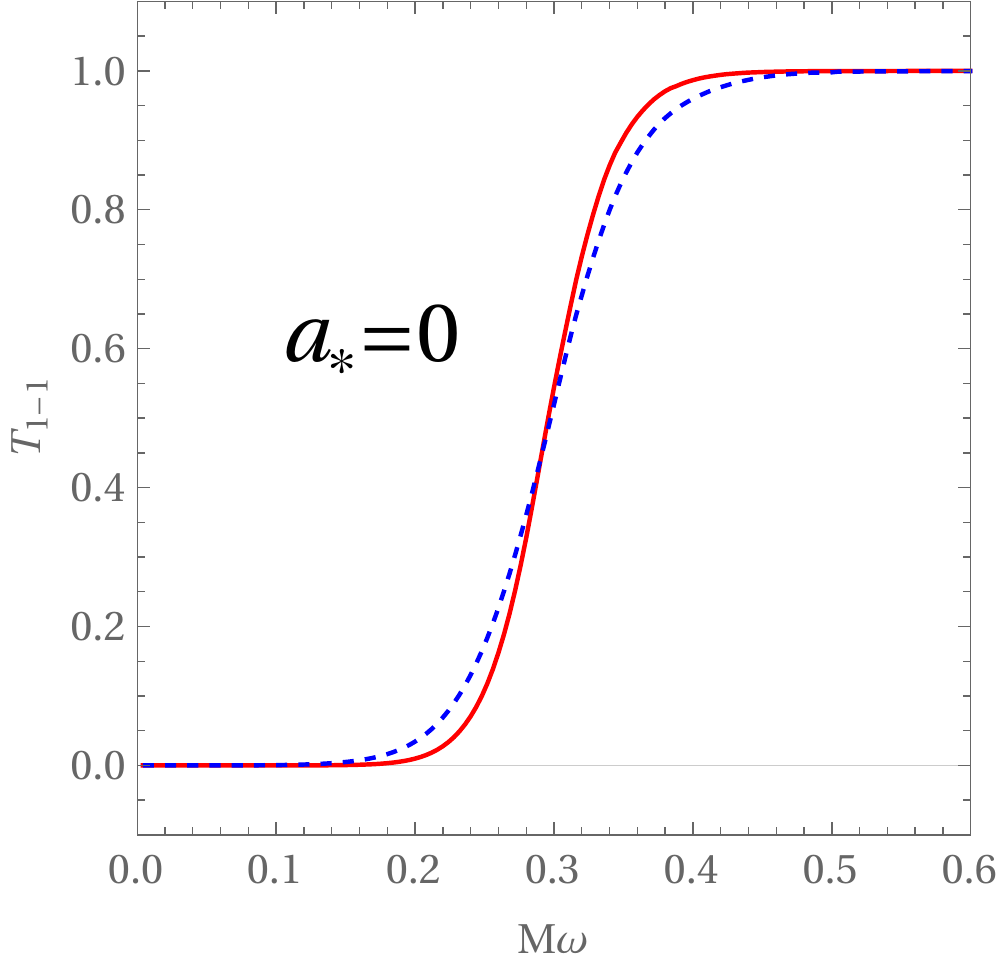}}
    \caption{GBFs of the mode $l=1 $, $m=-1$ of a BH rotating at $a_*=0.99$ in the case $\ell=0.99r_+$ (solid red line) and $\ell=0$ (blue dashed line).}
    \label{l=1 m=-1}
\end{figure}
\noindent The GBFs of the Kerr-black-bounce BH show a common behavior for the modes with $l\neq0$. When they are compared with the Kerr BH ones, they grow faster for frequencies lower than the main GBFs inflection point, on contrary, they grow slower for higher frequencies (as shown in Fig. \ref{l=1 m=-1} for the $l=1$, $m=-1$ mode). Also, this behavior is independent of the spin of the BH.\\
The scalar perturbation of both metrics show superradiant amplification if $\omega<m\Omega$. When this condition is met, both the GBFs have negative values, which are interpreted as wave amplification. Fig. \ref{T11} displays the comparison of the GBFs for the $l=m=1$ modes at $a_* =0.99$ highlighting the superradiant regime.\\
\begin{figure}[H]
    \centering
    \subfigure[]{\includegraphics[width=0.4\textwidth]{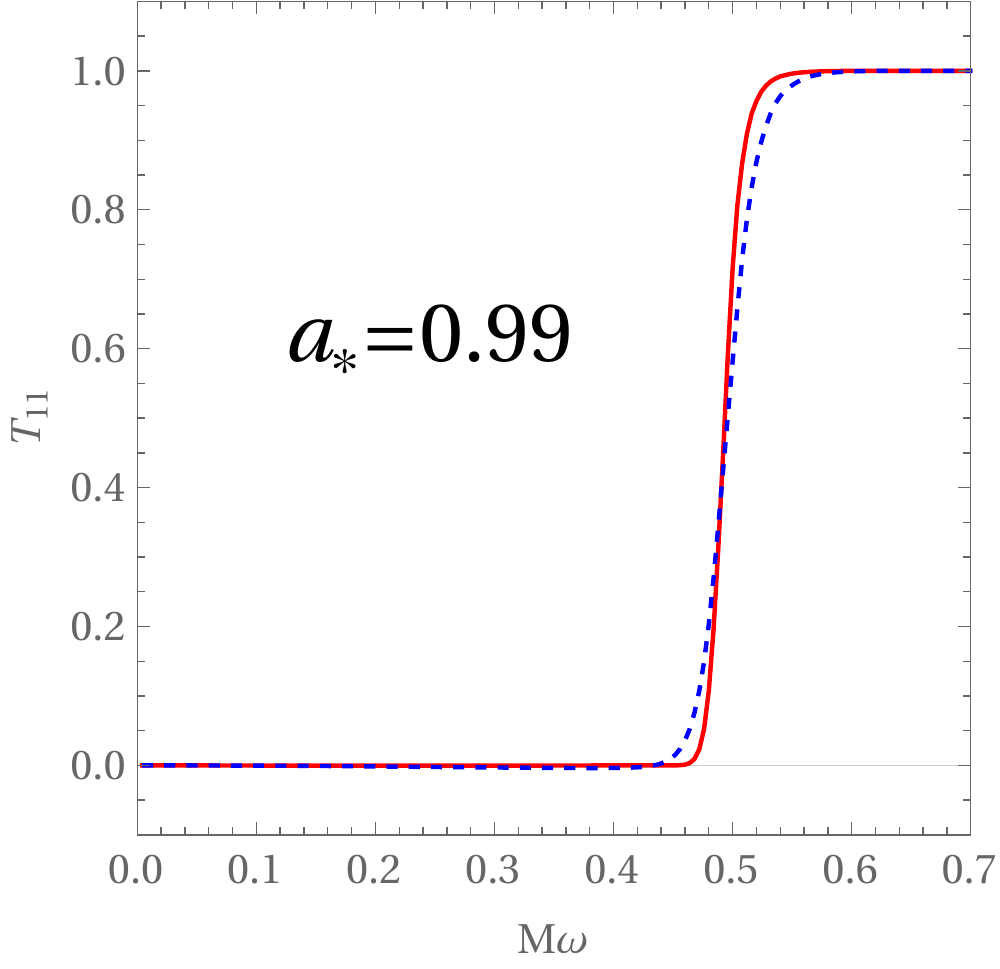}} 
    \subfigure[]{\includegraphics[width=0.4\textwidth]{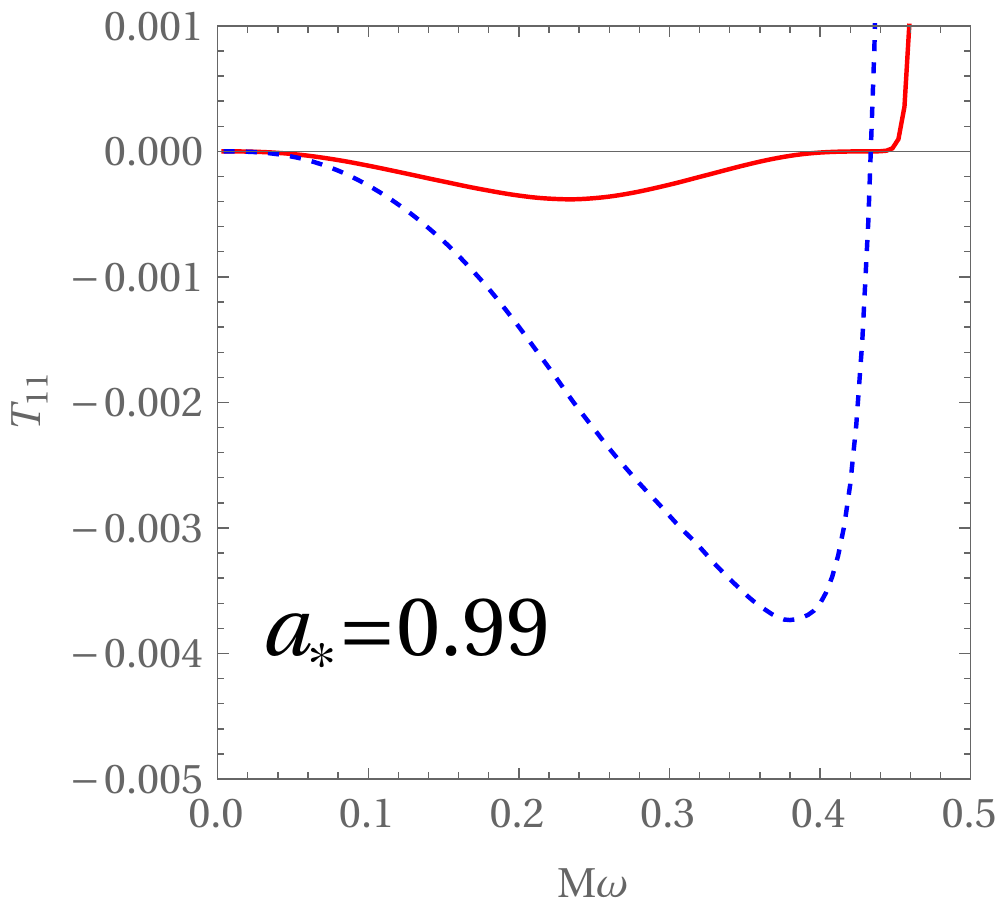}} 
    \caption{(a) Transmission coefficients of the mode $l=m=1$ of a BH rotating at $a_*=0.99$ in the case $\ell=0.99r_+$ (solid red line) and $\ell=0$ (blue dashed line). (b) Zoom of the superradiant regime of (a).}
    \label{T11}
\end{figure}
\noindent The Kerr-black-bounce GBFs show a less intense amplification and the shape of the superradiant regime peaks at lower frequencies. Also, the shape of the GBFs in the superradiant regime is different, being more symmetric than in the singular case. This result agrees with the tendency shown in the recent paper \cite{Franzin:2022iai}, in which it is reported that increasing the parameter $\ell$ causes a decrease in the superradiant amplification factor. Those are common features of all the superradiant modes. However, it has to be noticed that with an increasing azimuthal quantum number, the superradiant peak of the Kerr-black-bounce BH GBFs becomes smaller and smaller with respect to its singular counterpart.\\
\begin{figure}[H]
 \centering
    \subfigure[]{\includegraphics[width=0.4\textwidth]{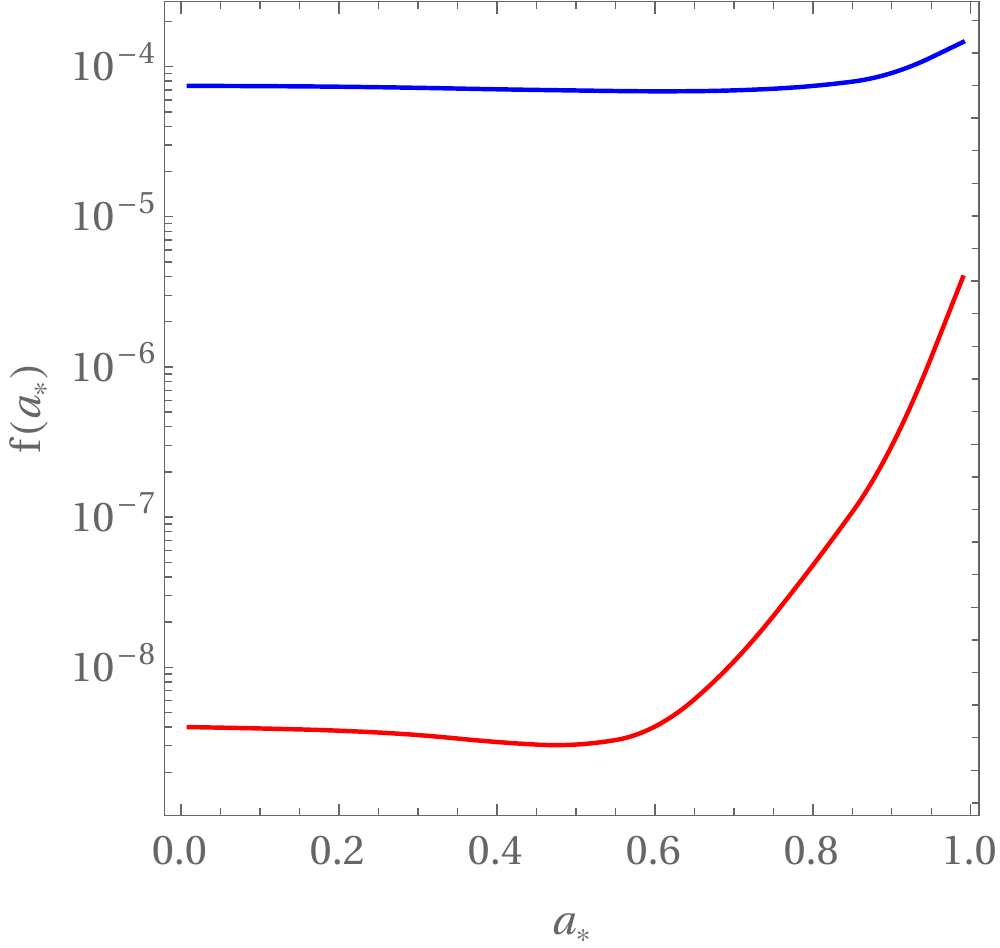}} 
    \subfigure[]{\includegraphics[width=0.4\textwidth]{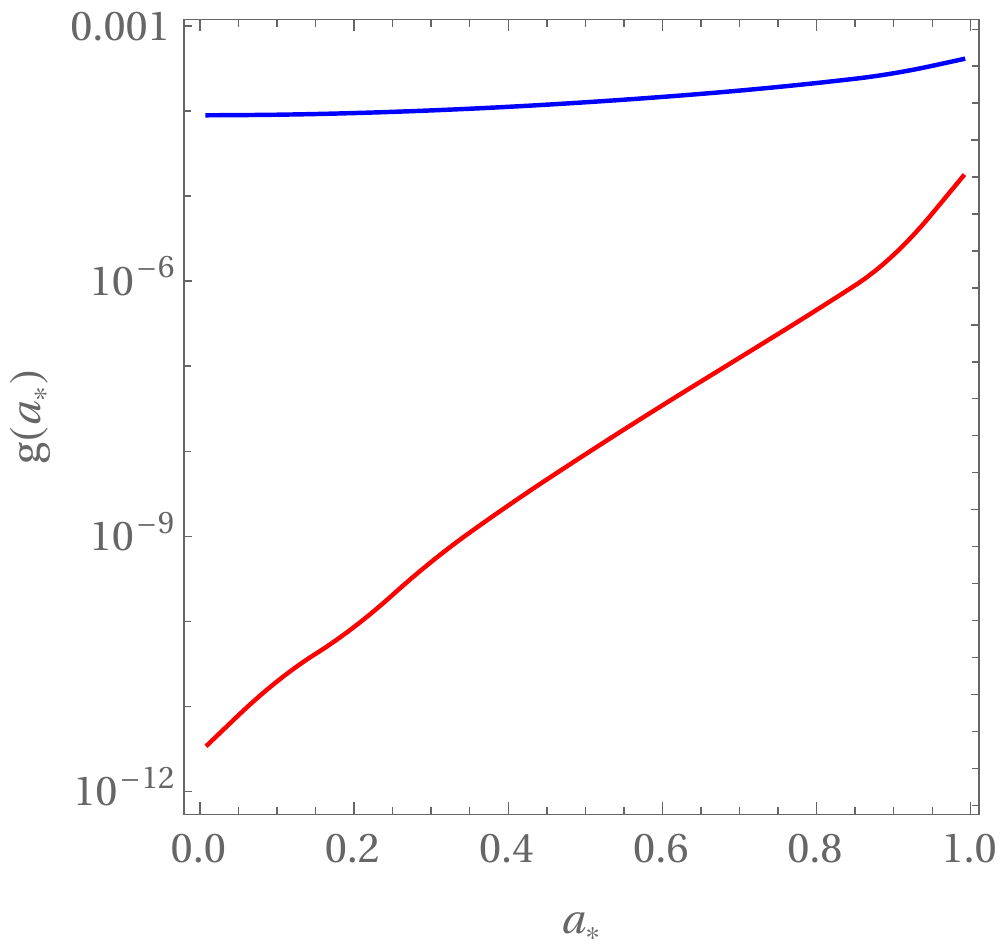}} 
    \caption{Plot of functions  f (a) and g (b) for different values of the BH spin parameter $a_*$. In the solid blue line is the Kerr BH, and in solid red the Kerr-black bounce}
    \label{fg}
\end{figure}
\noindent The functions $f$, and $g$, are calculated through (\ref{f_g}). The two BHs show different values of these functions. These are due to the above-mentioned GBF differences and in the different surface gravity (\ref{surface}), which plays a crucial role in the Bose-Einstein statistical factor in (\ref{f_g}) selecting lower frequency if $\ell\neq0$.
For these reasons, the Kerr-black-bounce BH functions $f$ and $g$ for $\ell=0.99 r_+$ are orders of magnitude smaller if compared with the singular case. Fig. \ref{fg} reports a comparison of those two cases.\\
For the same reasons, the functions $h$ are also different. It is shown in Fig.\ref{h} that the root of the Kerr BH is located at $\tilde{a}_{*}=0.555$, while the one of the Kerr-black-bounce is at $\tilde{a}_{*}=0.47$.\\
If the natal spin of both BHs is smaller than the respective root of $h$, the dominant emission mode is $l=0$. In this case, the evaporation due to a single scalar field will cause both BHs to lose mass faster than angular momentum. As a result, the evaporating BH will increase its value of $a_*$ up to the respective asymptotic value $\tilde{a}_{*}$. Conversely, the evolution of highly spinning BHs is dominated by higher $l$ modes decreasing the angular momentum of the BH and driving it toward its asymptotic values.\\
It is possible to speculate that the similar asymptotic value is due to the common origin of the gain/loss of dimensionless angular parameter. In fact, of the whole scalar modes emitted, $l=m=0$ solely does not subtract angular momentum and as it is reported in Fig.1 the transmission coefficients for the two analyzed BHs are the same for this mode. The differences are then related to the differences in the subdominant $l>0$ transmission coefficients. The dependency of the asymptotic BH spin value on the regularizing parameter is mild but present since a variation in the regularizing parameter incurs variations in the $l>0$ gray-body factors.

\begin{figure}[H]
 \centering{\includegraphics[width=0.4\textwidth]{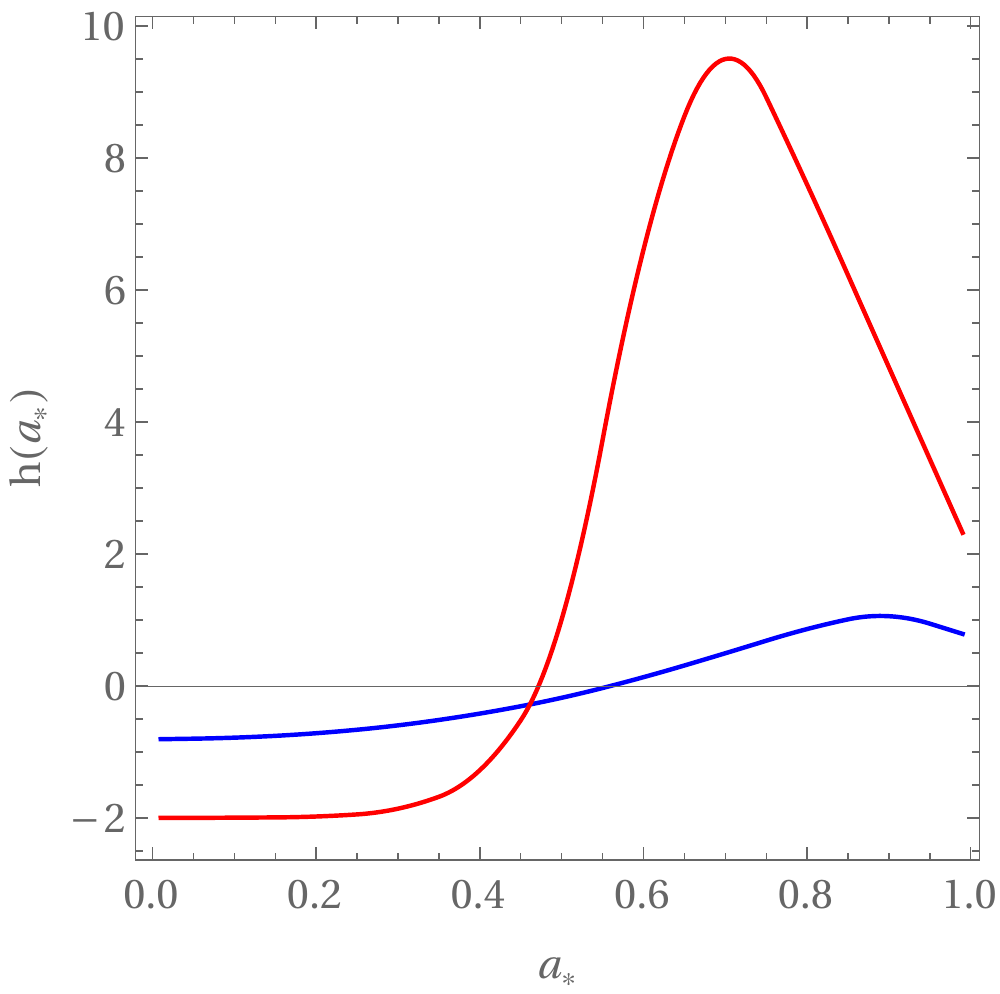}} 
    \caption{Plot of functions  $h=g/f - 2$  at different values of the BH spin parameter $a_*$. In the solid blue line is the Kerr BH, and in solid red the Kerr-black bounce}
    \label{h}
\end{figure}

\begin{figure}[H]
 \centering
    \subfigure[]{\includegraphics[width=0.33\textwidth]{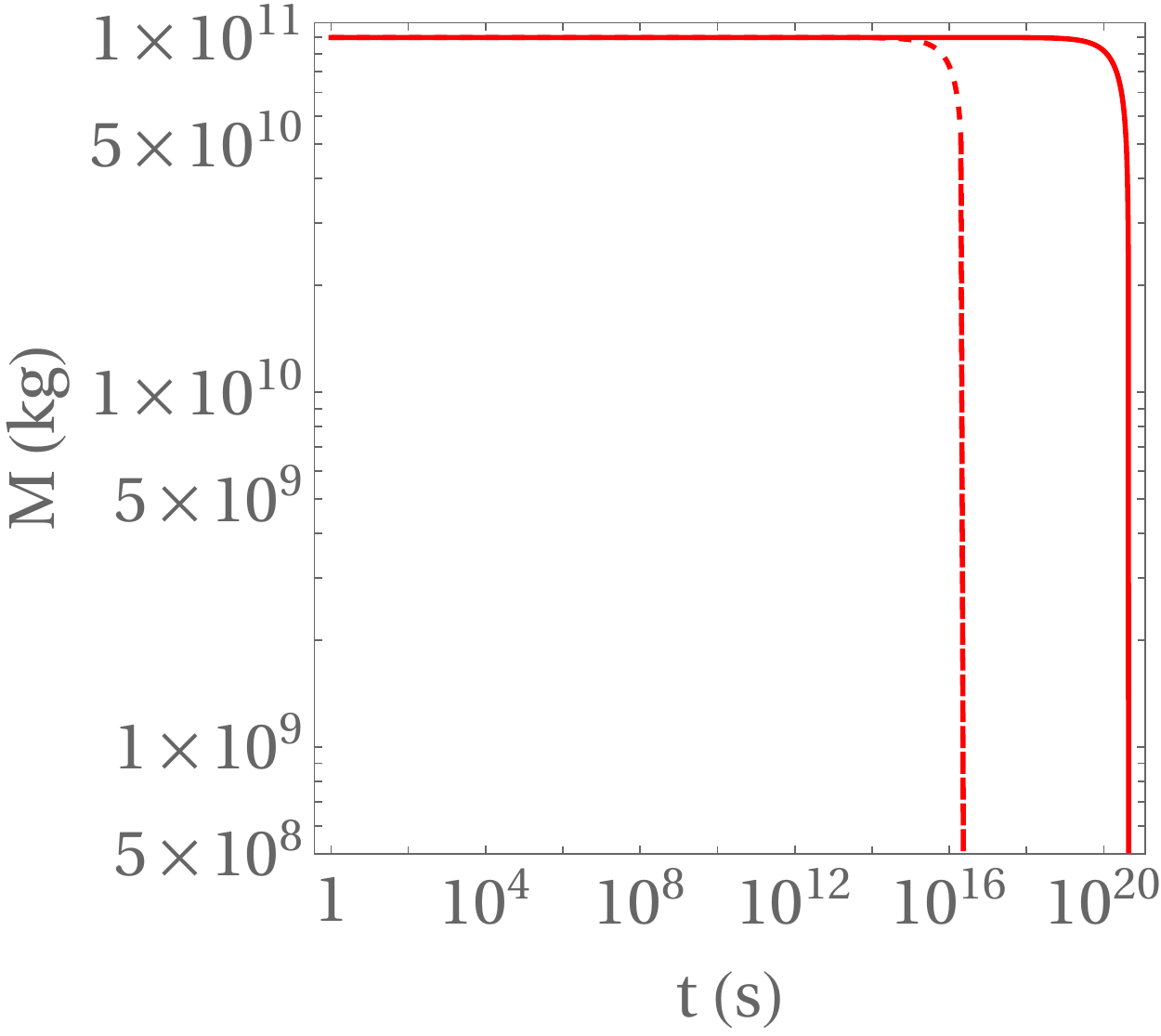}}
\end{figure}
\begin{figure}[H]
\centering
    \subfigure[]{\includegraphics[width=0.30\textwidth]{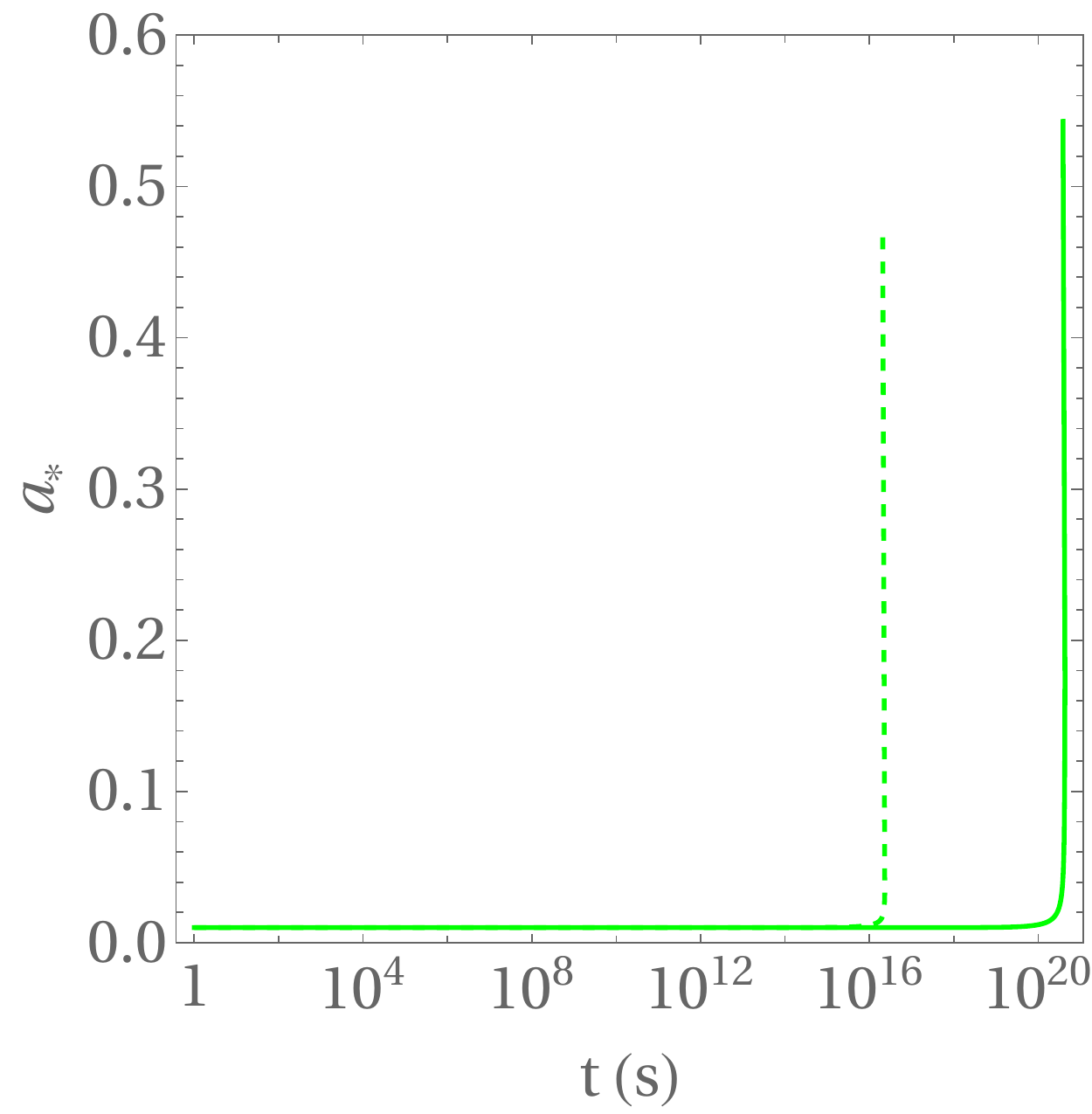}} 
\end{figure}
\begin{figure}[H]
    \hspace{1.11cm}\subfigure[]{\includegraphics[width=0.32\textwidth]{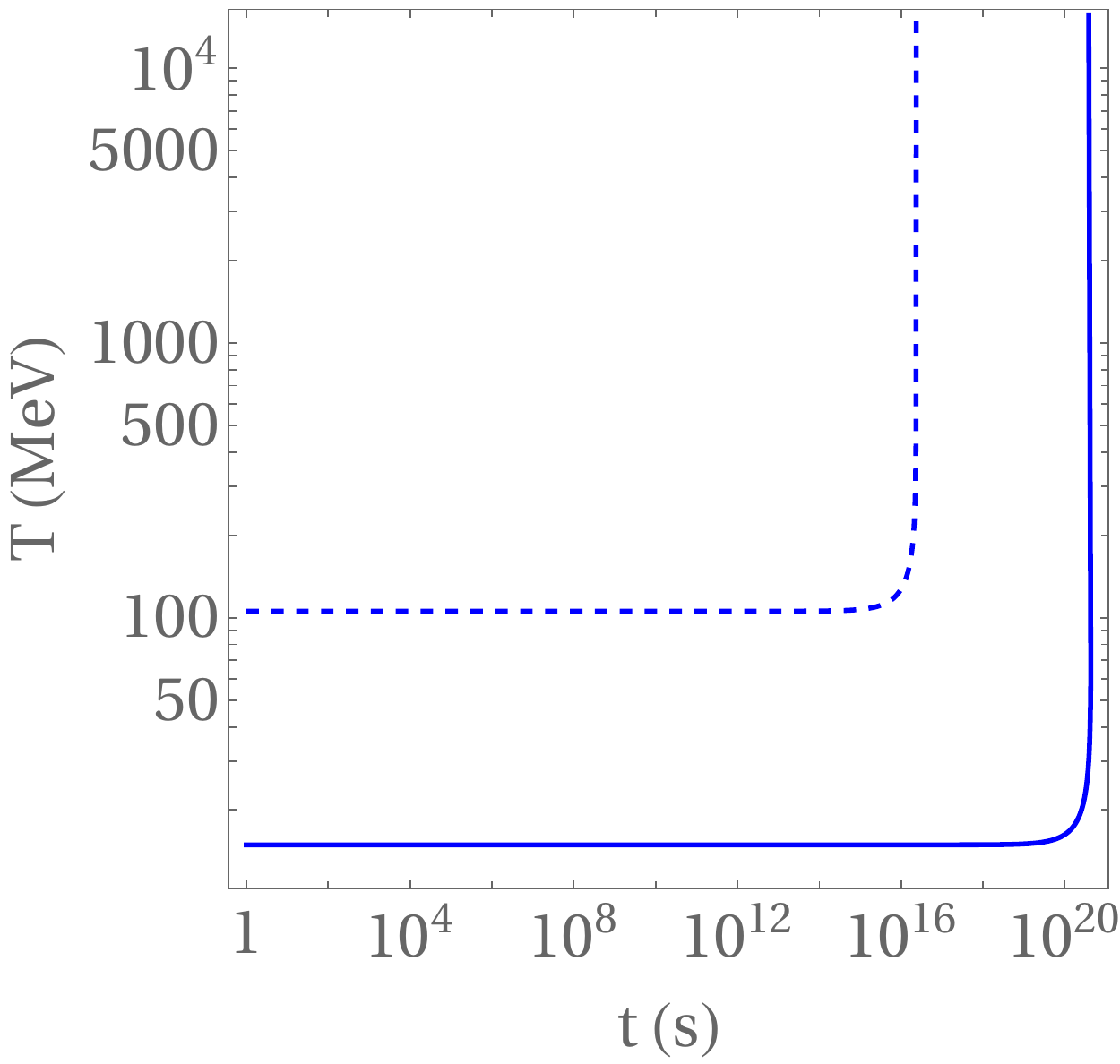}}
    \caption*{FIG. 6: Plot of the mass (a), spin (b), and temperature (c) as functions of the time, of a Kerr-black bounce having $\ell=0.99 r_+$ (solid lines) and a Kerr-black hole (dotted lines) of the same initial mass  $M_{K}=10^{11}$ kg, and spin parameter $a_{*i K}=0,01$, evolving by the emission of a single type of scalar particle.}
    \label{1}
\end{figure}

\begin{figure}[H]
   \hspace{1.4cm}\includegraphics[width=0.31\textwidth]{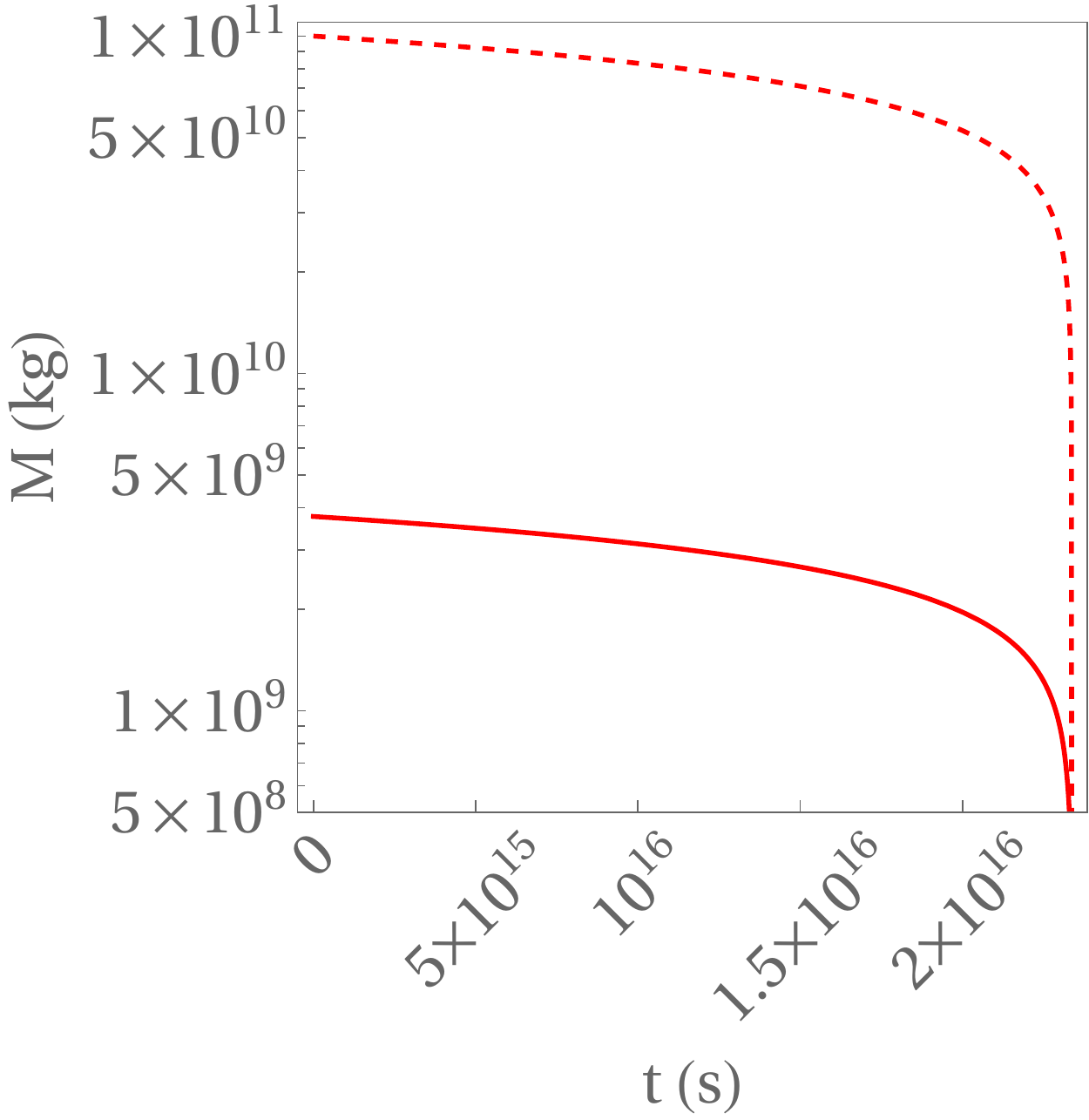}
   \caption*{\;\;\;\;\;\; {\footnotesize (a)}}
\end{figure}
\begin{figure}[H]\centering
   \includegraphics[width=0.27\textwidth]{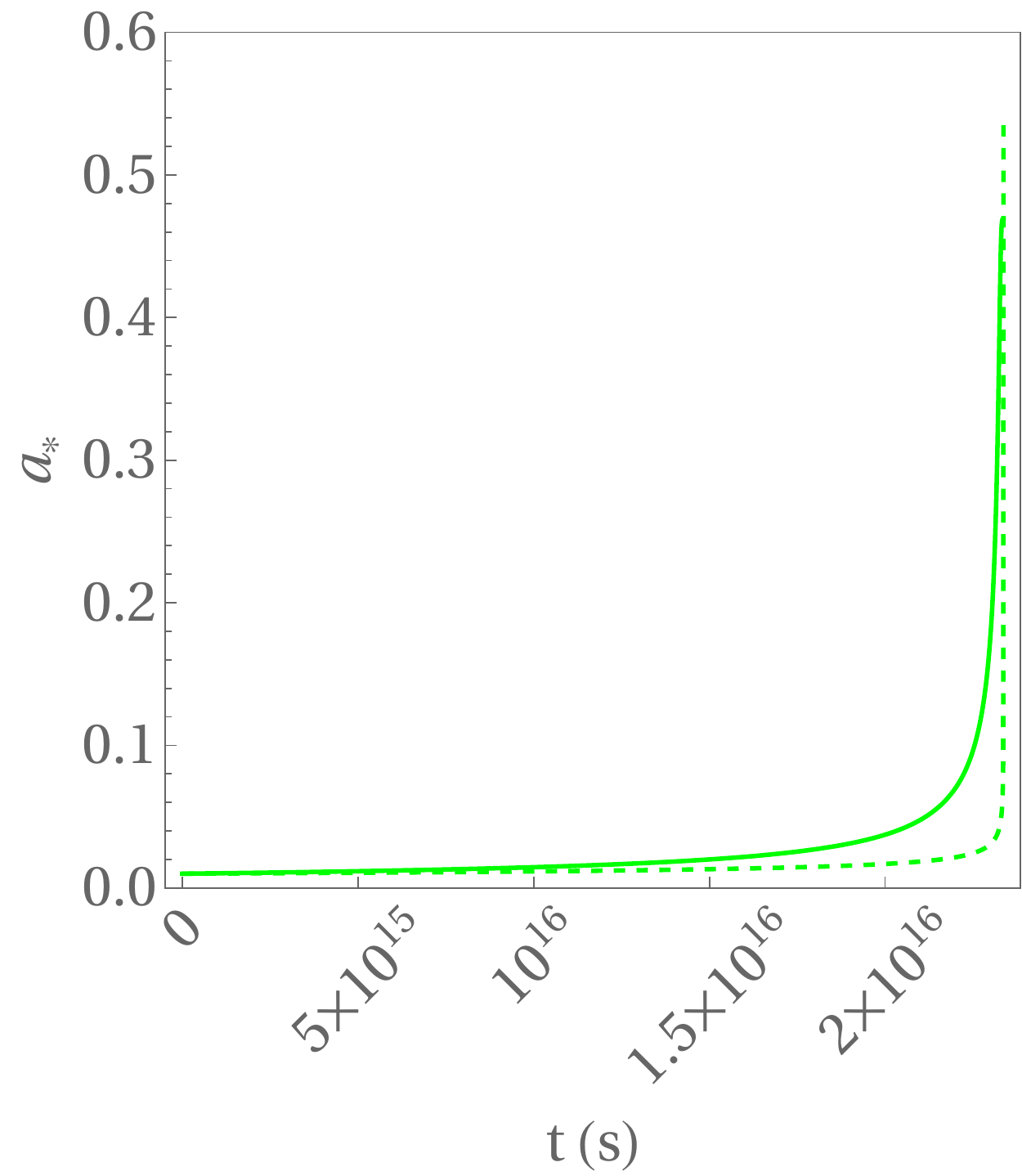}
   \caption*{\;\;\;\;\;\; {\footnotesize (b)}}
\end{figure}
\begin{figure}[H]
   \hspace{1.77cm}\includegraphics[width=0.29\textwidth]{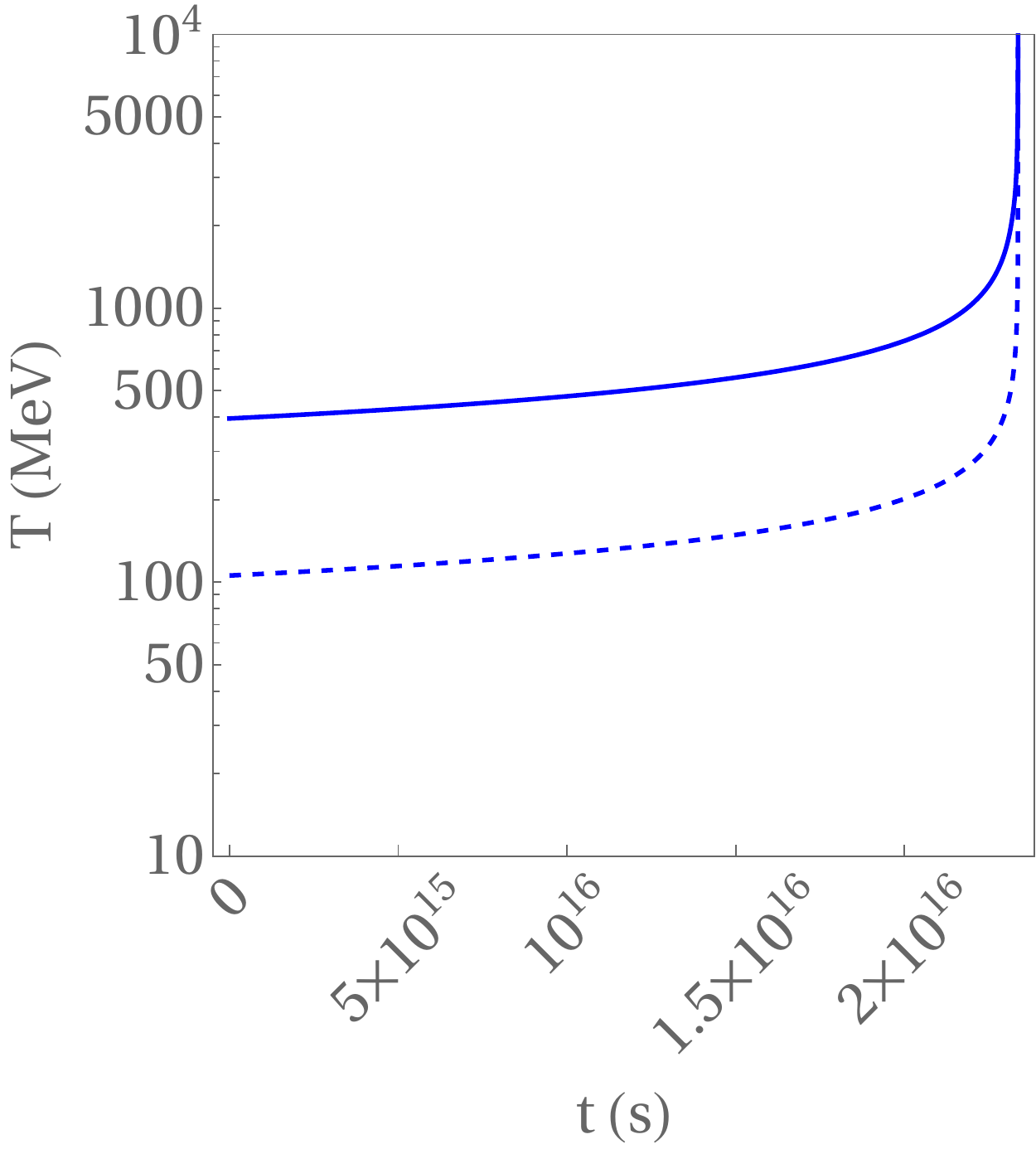}
   \caption*{\hfill {\footnotesize  (c)}\;\;\;\;\;\;\;\;\;\;\;\;\;\;\;\;\;\;\;\;\;\;\;\;\;\;\;\;\;\;\;\;\;\;\;\;\;\;\;\;  \\ \\ FIG. 7: Plot of the mass (a), spin (b), and temperature (c) as functions of the time, of the Kerr-black bounce having $\ell=0.99 r_+$ (solid lines) and the Kerr-black hole (dashed lines) of the same lifetime, evolving by the emission of a single type of scalar particle.}
\end{figure}
\noindent The regularizing parameter influencing the surface gravity plays a significant role in the dynamic evolution of the regular BH, which is much slower with respect to its singular counterpart. The lifetime of an isolated Kerr BH emitting only one scalar particle and having natal mass and spin of $M_{i}=10^{11}$ kg, and $a_{*i}=0,01$, is $\sim 2.34 \times 10^{16}$ s, while a nearly extremal Kerr-black-bounce BH with the same initial conditions has a lifetime of $\sim 4.37 \times 10^{20}$ s. Fig. 6 reports mass, spin parameter, and temperature as a function of time for such BHs.
It is interesting to consider two BHs of the same life span, and analyze their evolution. It is worth noticing that the time evolution of the spin parameters is different and the Kerr-black-bounce spin grows faster for most of the evolution as reported in Fig. 7.\\
Given its slower dynamical evolution, it is not surprising that the intensity peak of the primary emission for the regular BH is less intense with respect to a Kerr BH having the same mass and spin. This situation is reported in Fig. 8 (a) where masses of $M=3.5 \times 10^{10}$ kg and spin values of $a_*=0,0.9,0.99$ are considered. This plot shows a reduction in the number of emitted scalars as well as a reduction in the energy at which they are emitted, in line with the previous comments.
Finally, Fig. 8 (b) shows the primary emissivity for the same temperatures, namely, $301.93$, $183.35$, and $74.67$ MeV for $a_*=0,0.9,0.99$, respectively.\\
One may compare Fig. 8 with Fig. 2 of \cite{Arbey:2020yzj} which describes the primary emission of a Kerr BH for different field spins. Fig.2 of \cite{Arbey:2020yzj} highlights how the rotation in a Kerr BH reinforces the emission of non-spin-less particles and decreases the emission of scalar particles. This is no longer valid for the Kerr-bounce BH. In fact its scalar particle emissivity peaks at higher values for values of the spin parameter close extremality.
\begin{figure}[H]
 \centering
    \subfigure[]{\includegraphics[width=0.4\textwidth]{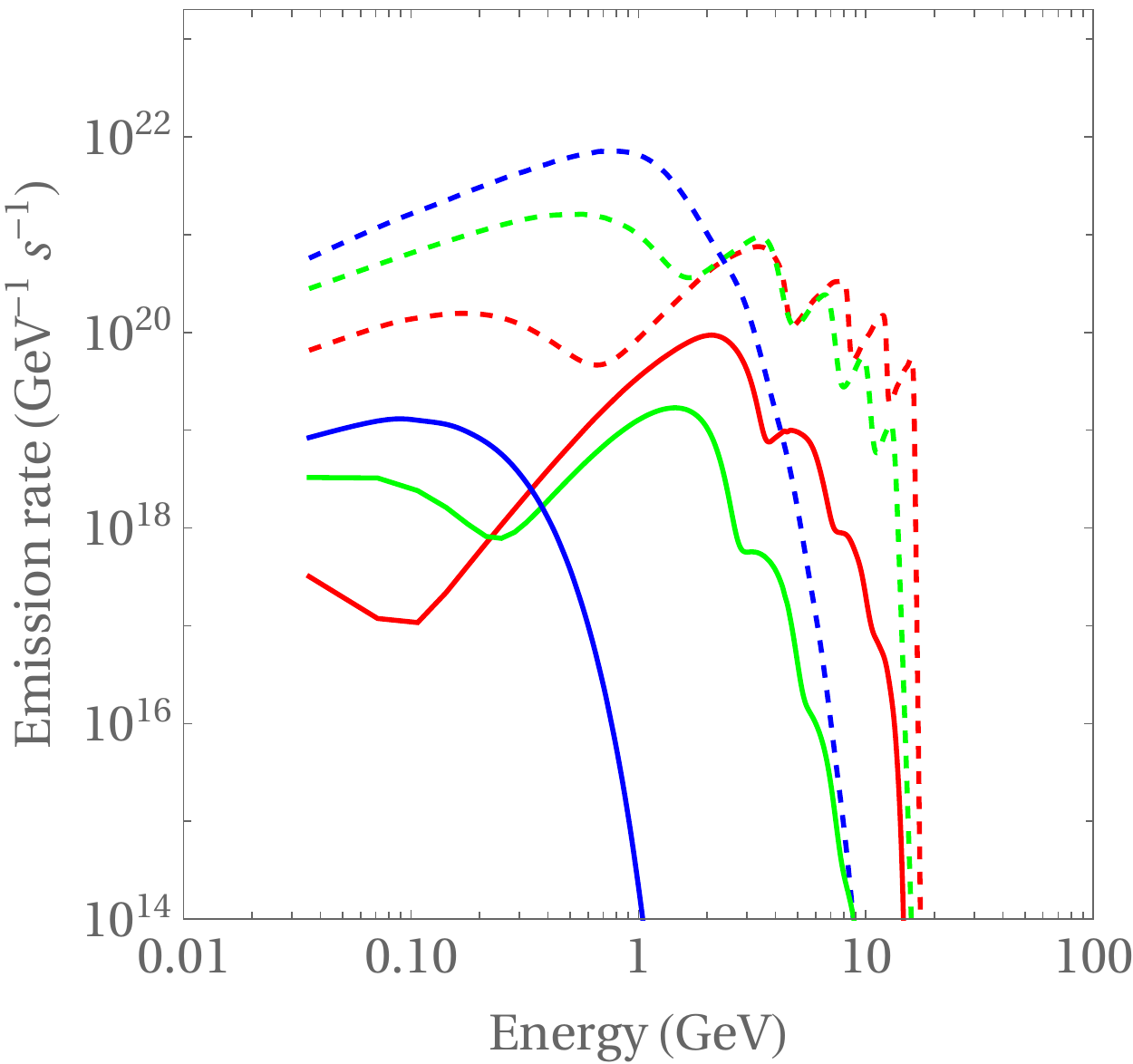}} 
    \subfigure[]{\includegraphics[width=0.4\textwidth]{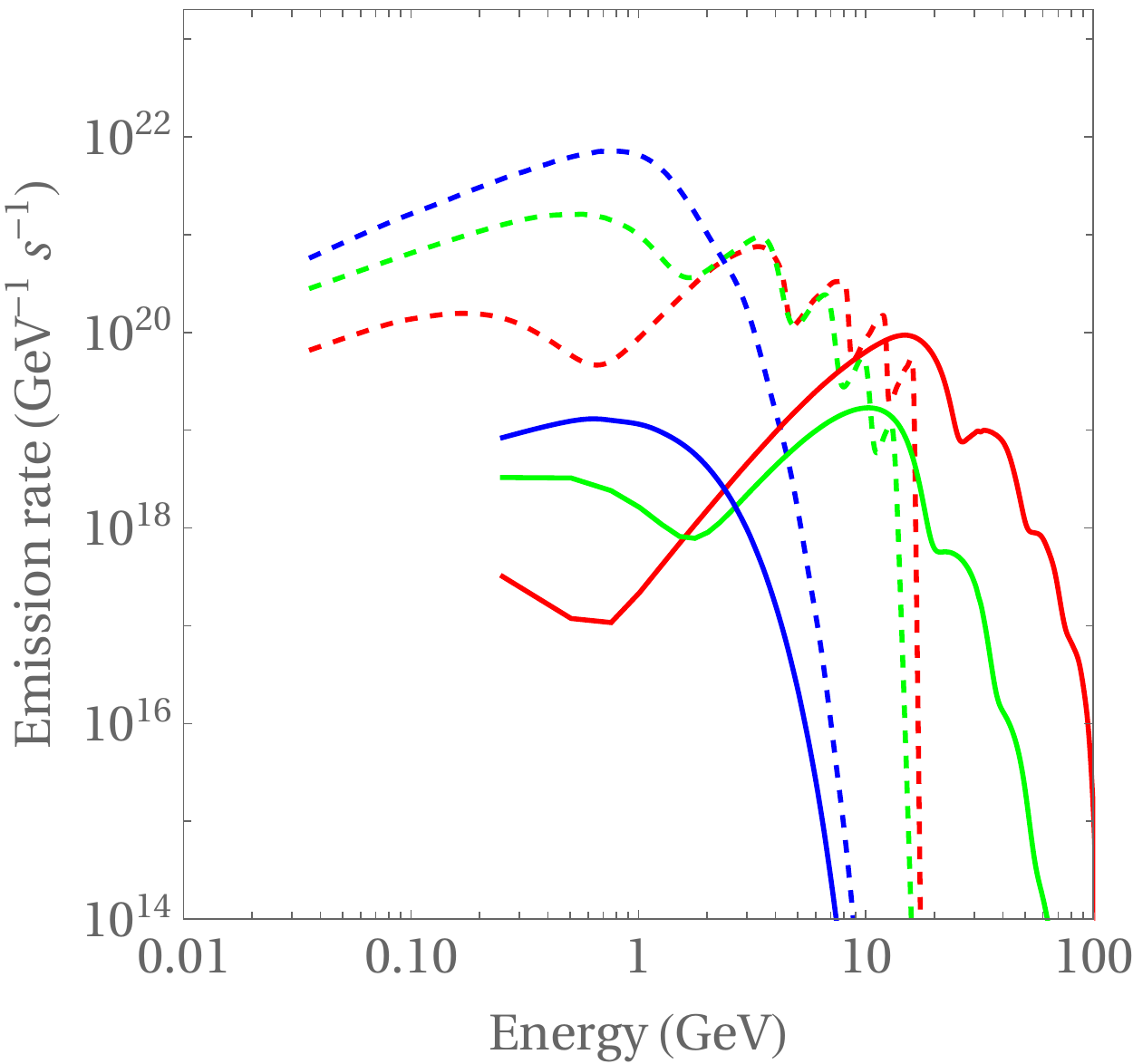}} 
    \caption*{FIG. 8: (a) Primary emissivity for regular (solid) and singular (dashed) BHs in the case of same masses of $M=3.5 \times 10^{10}$ kg for spin values of $a_*=0,0.9,0.99$ in blue, green, and red, respectively. (b) Primary emissivity for the same BHs in the case of the same temperature, namely $301.93$ MeV for $a_*=0$ in blue, $183.35$ MeV for $a_*=0.9$ in green, and $74.67$ MeV for $a_*=0.99$ in red.}
    \label{emiss}
\end{figure}
 \vspace{3cm}

\section{Conclusions}

In this paper, we studied the evolution, under the emission of scalar radiation via the Hawking process, of a rotating regular black hole described by the Kerr-black-bounce metric. The study is performed in the case of a nearly extremal value of the regularizing parameter ($\ell = 0.99 r_+$). The differences in the dynamics of the evaporation of such BH and a Kerr BH are outlined. Namely, the negative transmission coefficients regime, the asymptotic value of $a_*$, the emissivity, and the lifetime are discussed and compared.\\ 
The main lesson of this toy-model points towards a possible investigation of beyond-the-horizon features by analyzing the Hawking radiation. For example, by assuming a way to infer the BH mass and spin independently from the primary Hawking emission, it is possible by measuring the peak intensity to obtain an indirect measure of $\ell$ in the contest of the Kerr-black-bounce solutions, and in general, would provide a measure of how much the BH solution differs from the Kerr one. It is most likely to observe the Hawking emission of photons and not scalar particles, but, since the definition of $f$ and $g$ for spin-1 bosons is given by Eq. (\ref{f_g}) with the appropriate GBFs, one can expect that the differences between the Kerr solution and the regular one are still present. This work also suggests that tracking the time evolution of the spin parameter could provide information on the spacetime structure.\\
Such characteristics are certainly irrelevant for BHs of the size measured today but may become a powerful and handy tool in light of possible future primordial BHs detection.\\
We leave GBFs calculation for spin 1/2, 1, and 2 fields and implementation of an accurate evaporation scenario for future studies.\\
In a standard evolution scenario, BHs clearly do not evaporate through the sole emission of a scalar field. Nevertheless, scenarios involving the conspicuous presence of scalar particles such as the string axiverse \cite{Calza:2021czr,Arvanitaki:2009fg} may display similar characteristics. In fact, in the limit of many axion-like particles, the emission of scalar particles dominates the evolution which results similar, up to a normalization, to the single scalar case.

\newpage

\section*{Acknowledgements}
I would like to thank  Violetta Sagun, Massimiliano Rinaldi, and Jo\~ao Rosa, for the fruitful discussions we had. \newline
This work was supported by national funds from FCT, I.P., within the projects UIDB/04564/2020, UIDP/04564/2020 and the FCT-CERN project CERN/FIS-PAR/0027/2021. \newline
M.C. is also supported by the FCT doctoral grant SFRH/BD/146700/2019.

\end{document}